\documentclass[journal]{IEEEtran}
\ifCLASSINFOpdf
\else
\fi
\usepackage[cmex10]{amsmath}
\usepackage{algorithmic}
\usepackage{array}
\usepackage{mdwmath}
\usepackage{mdwtab}
\usepackage{eqparbox}
\usepackage{graphicx}
\usepackage[tight,footnotesize]{subfigure}
\usepackage[caption=false,font=footnotesize]{subfig}
\usepackage{fixltx2e}
\usepackage{stfloats}
\fnbelowfloat
\ifCLASSOPTIONcaptionsoff
  \usepackage[nomarkers]{endfloat}
 \let\MYoriglatexcaption\caption
 \renewcommand{\caption}[2][\relax]{\MYoriglatexcaption[#2]{#2}}
\fi
 \let\MYorigsubfloat\subfloat
 \renewcommand{\subfloat}[2][\relax]{\MYorigsubfloat[]{#2}}
 \let\MYorigsubfigure\subfigure
 \renewcommand{\subfigure}[2][\relax]{\MYorigsubfigure[]{#2}}
\usepackage{url}
\begin{document}
\title{Modelling and Control of Quantum Measurement Induced Backaction in Double Quantum Dots}

\author{Wei~Cui,~\IEEEmembership{Senior Member,~IEEE}
        and~Daoyi~Dong,~\IEEEmembership{Senior Member,~IEEE}

\thanks{This work was supported by the National Natural Science Foundation of China under Grant 11404113, the Australian Research Council's Discovery projects funding scheme under Project DP130101658,
 and the Guangzhou Key Laboratory of Brain Computer Interaction and Applications under Grant 201509010006.
}
\thanks{W. Cui is with the School of Automation Science and Engineering, South China University of Technology, Guangzhou 510641, China (e-mail: aucuiwei@scut.edu.cn).}
\thanks{D. Dong is with the School of Engineering and Information Technology, University of New South Wales, Canberra, ACT 2600, Australia (e-mail:  daoyidong@gmail.com).}
}

\markboth{Submitted to IEEE Transaction on Control Systems Technology}%
{Shell \MakeLowercase{\textit{et al.}}: Bare Demo of IEEEtran.cls for Journals}
\maketitle

\begin{abstract}
Quantum measurements disturb the quantum system being measured, and this is known as measurement-induced backaction. In this work, we consider a double quantum dot monitored by a nearby quantum point contact where the measurement-induced backaction plays an important role. Taking advantage of the quantum master equation approach, we calculate the tunnelling current, and propose a simple feedback-control law to realize and stabilize the tunnelling current. Theoretical analysis and numerical simulations show that the feedback control law can make the current quickly convergent to the desired value.
\end{abstract}


\IEEEpeerreviewmaketitle

\section{Introduction}

\IEEEPARstart{C}{o}ntrolling quantum systems is an essential task in many applications including quantum information, atomic physics and molecular chemistry \cite{book1,DongSurvey,DongICST,WuTCST,Zhang et al 2012,ZhangICST}. In contrast to classical systems, where measurements do not alter the state of the system, quantum projective measurement \cite{Neumann1955,Kofman2012,Kurt2014} collapses a quantum system  into one of its eigenstates in a probabilistic manner: the ``measurement-induced backaction" (see, e.g., \cite{Dressel2014,Devoret2013,Cui2012,Kulkarni2014} and references therein).
Recent progress in macroscopic quantum manipulation and quantum weak measurements have made it possible to perform experiments in which individual quantum systems can be monitored and feedback-controlled in real time   \cite{Mirr2011,Furu2012,Iida2012,Zhang2014}.
 In this paper, we derive a non-Markovian master equation  for a double quantum dot (DQD) \cite{Mi2017} monitored by a nearby quantum point contact (QPC) where the measurement-induced backaction plays a significant role. In this master equation, time-dependent coefficients can characterize non-Markovian effects.
Furthermore,  we consider how to reduce the measurement-induced backaction and improve the quantum measurement efficiency by using quantum control theory \cite{Dong:2012, James2010,Guo2013,Yamamoto:2009}.

Several strategies \cite{Wiseman1995,Bran2007} have already been proposed to reduce quantum measurement-induced backaction. For example,
 Wiseman \cite{Wiseman1995} considered the measurement-induced backaction based on quantum trajectory theory. In the case of damping, the jump operator is not Hermitian, and so the measurements permitted by damping are necessarily quantum-demolition measurements. He  proved that by feeding back the measurement result to control the system dynamics, the quantum-demolition measurement can be turned into a quantum-nondemolition measurement \cite{Wiseman1995}. The scheme to reduce and suppress the measurement-induced backaction is to obtain a feedback Hamiltonian based on the measurements on the system.
A crucial issue  in this approach is the necessity to solve a non-linear  equation to find the feedback control law.
The authors in \cite{Bran2007} used measurement and feedback control to attempt to correct the state of the qubit.  They demonstrated that projective measurements are not optimal for this task, and that there exists a nonprojective measurement with an optimal measurement strength which achieves the best trade-off between gaining information about the system and disturbing it through measurement backaction. Similar to \cite{Wiseman1995}, the feedback control is a unitary rotation whose angle depends on the measurement result.

DQD has been considered as a promising candidate of solid state qubits for  quantum computation.
This system  can be monitored  by coupling it to a nearby biased quantum point contact. It is interesting to understand the backaction disturbance resulting from such a detection approach. For a zero-bias DQD, the effect of charge-detector-induced backaction was  theoretically studied to explain experimental observations of inelastic electron tunneling \cite{You2010,Li2012}.
 Ref.~\cite{You20132} exposed that under certain conditions the QPC-induced backaction has profound effects on the counting statistics, e.g., changing the shot noise from being sub-Poissonian to super-Poissonian, and changing the skewness from  positive to negative. However, how to effectively reduce this measurement-induced backaction is still an open question.
Two possible ways have been proposed to reduce the backaction effects by making use of this interference: 1) turn the DQD to an operating point where destructive interference suppresses phonon absorption, and 2) manipulate the electron-phonon interaction. There are various limitations on the applications of these existing strategies \cite{Wiseman1995}. First, it is difficult to estimate quantum states and the measurement-induced backaction from incomplete measurement data. Second, the system of interest is generally open, which means that the dynamics of its environment will affect its evolution. Thus, an accurate master equation is needed to characterize the whole measurement model.

In this paper, a non-Markovian master equation for a DQD monitored by a nearby QPC is derived \cite{CuiCCC}. Based on an ensemble description of the system of interest, we consider the quantum detection over many identical systems simultaneously.
Moreover, we develop a simple feedback control strategy based on counting statistics to estimate and compensate the measurement-induced backaction in the DQD.

This paper is organized as follows. In Section~II we describe a model of a double quantum dot measured by a nearby quantum point contact.  Using the quantum master equation, the tunnelling current is calculated in Section~III. In Section~IV, we design a feedback controller to realize and stabilize a target tunnelling current. Conclusions are given in Section~V.

\section{Physical System}
DQD has been considered as a promising candidate to prepare solid state qubits for quantum computation \cite{Chen2017,Stockklauser2017,Schuetz2017}.
This system can be monitored by a nearby QPC and controlled by electronic signals or picosecond laser pulses \cite{Brandes2010,White2010}. The measurement mechanism can be understood as follows. QPC measurements rely on the Coulomb interaction between the DQD and the QPC. The current in the QPC is sensitive to the charge state of the DQD. By measuring the current passing through the QPC one can infer the quantum state of the DQD. However, the measurement-induced backaction will collapse the quantum state. Theoretical  and experimental  results demonstrated that the backaction of the QPC on the DQD leads to a quantum limit of detection and decoherence \cite{Li2005}.

We consider a DQD in the Coulomb-blockade regime with strong intradot and interdot interaction (see Fig. 1(a)). The Hilbert space of the DQD system spanned by three states $|0\rangle,|1\rangle$, and $|2\rangle$.
Here, the two basis states $|1\rangle\equiv|1,0\rangle$ and $|2\rangle\equiv|0,1\rangle$ describe, respectively, one additional electron in the left and right dot above the empty state $|0\rangle\equiv|0,0\rangle$. Accordingly, the current through the QPC  switches between these three different values. The total Hamiltonian of the DQD system is
\begin{eqnarray}
H_{0}=H_{{\rm DQD}}+H_{{\rm Leads}}+H_{{\rm T}},
\end{eqnarray}
where the dynamics of the interdot DQD system is described by the Hamiltonian ($\hbar=1$),
\begin{eqnarray}
H_{{\rm DQD}}=\epsilon_{1}d_{1}^{\dag}d_{1}+\epsilon_{2}d_{2}^{\dag}d_{2}+\Omega(d_{1}^{\dag}d_{2}+d_{2}^{\dag}d_{1}).
\end{eqnarray}
Here, $\epsilon_{i}~(i=1,2)$ are the energies for a single electron state in each dot, $d_i^{\dag}$ is the transpose and conjugate of $d_i$,
$\Omega$ is the coupling strength between the two dots, and $d_{i}~(d_{i}^{\dag})$ represents the electron annihilation
(creation) operator in each dot. This system can be approximated as
a two-level system, and it is characterized by the energy offset,
$\epsilon=\epsilon_{2}-\epsilon_{1}$. The effective Hamiltonian can be written as
\begin{eqnarray}\label{Ham}
H_{{\rm DQD}}=\frac{\epsilon}{2}\sigma_{z}+\Omega\sigma_{x}
\end{eqnarray}
where $\sigma_{z}=d_{2}^{\dag}d_{2}-d_{1}^{\dag}d_{1}$ and $\sigma_{x}=d_{2}^{\dag}d_{1}+d_{1}^{\dag}d_{2}$
are the Pauli matrices for the pseudospin bases of $|1\rangle$ and $|2\rangle$.
By diagonalizing the DQD Hamiltonian, we can rewrite it as
\begin{eqnarray}
H_{{\rm DQD}}=\omega_{0}(|e\rangle\langle e|-|g\rangle\langle g|)/2=\omega_{0}\varrho_{z}/2,
\end{eqnarray}
where $\varrho_{z}=|e\rangle\langle e|-|g\rangle\langle g|$, and
the eigenstates are $$|g\rangle=\alpha|1\rangle-\beta|2\rangle,$$
$$|e\rangle=\beta|1\rangle+\alpha|2\rangle,$$ with $\alpha\equiv\cos(\theta/2),$ and $\beta\equiv\sin(\theta/2)$.
Here $\theta$ is given by $\tan\theta=2\Omega/\epsilon$. The coherent
oscillations of the system have a frequency
$\omega_{0}=\sqrt{\epsilon^{2}+4|\Omega|^{2}}$.

The Hamiltonian of the two electrodes in the DQD is
\begin{eqnarray}
H_{{\rm Leads}}=\sum_{k}\epsilon_{{ l}k}c_{{ l}k}^{\dag}c_{{{ l}k}}+\epsilon_{{ r}k}c_{{ r}k}^{\dag}c_{{ r}k},
\end{eqnarray}
where $c_{{ l}k}~(c_{{ r}k})$ is the annihilation operator
of an electron in the left (right) lead with quantum numbers $k$, and $\epsilon_{{ l}k}~(\epsilon_{{ r}k})$ is the energy of the annihilation operator $c_{{ l}k}~(c_{{ r}k})$. The
tunneling Hamiltonian of the DQD and the two electrodes is
\begin{eqnarray}
H_{{\rm T}}=\sum_{k}\bigg(\Omega_{{l}k}c_{{l}k}d_{1}^{\dag}+\Omega_{{r}k}c_{{r}k}^{\dag}\Lambda_{{r}}^{\dag}d_{2}\\\nonumber+\Omega_{{l}k}^{\ast}c_{{l}k}^{\dag}d_{1}+\Omega_{{r}k}c_{{r}k}\Lambda_{{r}}d_{2}^{\dag}\bigg),\label{interaction1}
\end{eqnarray}
which depends on the tunneling coupling strengths $\Omega_{{l}k}$
and $\Omega_{{r}k}$. Here, $\Lambda_{{r}}^{\dag}$ is the operator to count the number of tunnelled electrons \cite{You2010}. The behavior of the DQD tunnelling process is schematically depicted in Fig.~1(b).

\begin{figure}
\setlength{\abovecaptionskip}{6pt}
\centerline{\scalebox{1.2}[1.2]
{\includegraphics{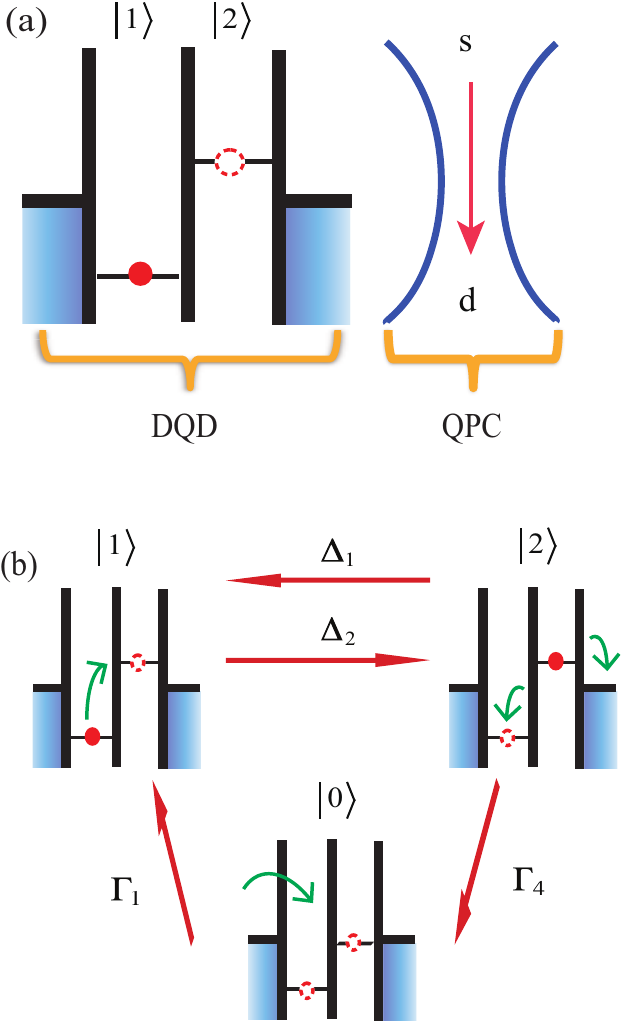}}}
\caption{(a) Schematic diagram of a double quantum dot (DQD) measured by a nearby quantum point contact.
The s and d in the QPC denote the source and the drain.
(b) Energy-level diagrams for the three states of the DQD. The arrows represent the  directions of the electron transitions. The tunnelling rates $\Gamma_{1,4}$ are defined in Eq. (\ref{Gamma}).
The decay rates $\Delta_{1,2}$ are defined in Eq. (\ref{Delta1}).
}
\end{figure}

The DQD is measured by a low-transparency QPC, and the total Hamiltonian of this composite system is
\begin{eqnarray}
H_{{\rm tot}}=H_{0}+H_{{\rm QPC}}+H_{{\rm int}},
\end{eqnarray}
where the Hamiltonians of the QPC and the DQD-QPC interaction are
\begin{eqnarray}
H_{{\rm QPC}}=\sum_{m}\epsilon_{{s}m}a_{{s}m}^{\dag}a_{{s}m}+\sum_{n}\epsilon_{{d}n}a_{{d}n}^{\dag}a_{{d}n},
\end{eqnarray}
\begin{eqnarray}
H_{{\rm int}}=\sum_{m,n}(D-\chi_{1}^ {}d_{1}^{\dag}d_{1}-\chi_{2}^ {}d_{2}^{\dag}d_{2})(a_{{s}m}^{\dag}a_{{d}n}+a_{{d}n}^{\dag}a_{{s}m}).\label{interaction12}
\end{eqnarray}
Here $a_{{s}m}~(a_{{d}n})$ is the electron annihilation operator for electrons in the source (drain) of the QPC with momentum $m\ (n)$, and $\epsilon_{{s}m}~(\epsilon_{{d}n})$ is the corresponding energy.
When a QPC is placed near a DQD, it can be used as an ideal detector to implement quantum weak measurements on the DQD, where the tunnelling barrier of the QPC is modulated by the charge in the nearby DQD. In Eq.~(\ref{interaction12}), $D$ is the tunnelling amplitude of an isolated QPC, $\chi_{1}^ {}~(\chi_{2}^ {})$ provides the variation in the tunnelling amplitude when the extra electron stays at the first (second) dot. We assume $\chi_{1}<\chi_{2}$, because the QPC is located closer to the second dot. The current through the QPC is sensitive to the electron location and switches between three values corresponding to the three charge states of the DQD. Figure 1(b) depicts the electron tunnelling in different directions. The current in the QPC allows to obtain  quantum state information on the DQD \cite{Poltl2011}.

\section{Non-Markovian master equation}

Quantum master equation approach \cite{Breuer2002} provides a theoretical tool to understanding the dynamics of open quantum systems.
In the high-bias case, the behavior of the system density matrix can be described with a master equation of the Lindblad form \cite{You2010,You20132,Li2005,Gurvitz1996}.
In the interaction picture with respect to $$H^{\prime}=H_{{\rm DQD}}+H_{{\rm QPC}},$$ the quantum system is governed by
\begin{eqnarray}
\dot{\rho}_{s}(t)=-\int_{0}^{t}{\rm d}t^{\prime}{\rm Tr}_{{\rm E}}\bigg[\big[\rho_{{ s}}(t)\otimes\rho_{{\rm E}},H_{{\rm I}}(t^{\prime})\big],H_{{\rm I}}(t)\bigg], \label{Master equation}
\end{eqnarray}
where ${\rm Tr}_{{\rm E}}$ indicates the trace over the environmental degrees of freedom, $\rho_{E}$ is the state of the environment and $\rho_{s}$ is the system state. The commutator $[\cdot~,~\cdot]$ is defined by $[A,B]=AB-BA$.
The DQD interacts with two environments: the QPC and two electrodes. Thus, the interaction Hamiltonian is \cite{You2010}
\begin{eqnarray}\label{Hamiltonian}
H_{{\rm I}}(t)=A(t)Y(t)+H_{{\rm T}}(t).
\end{eqnarray}
The first term of the right hand side of Eq. (\ref{Hamiltonian}) is the interaction with the QPC and
\begin{eqnarray}
A(t) & = &e^{iH_{\rm DQD}t}\big(D-\chi_{1}^ {}d_{1}^{\dag}d_{1}-\chi_{2}^ {}d_{2}^{\dag}d_{2}\big)e^{-iH_{\rm DQD}t}\nonumber\\
      & = &\sum_{j=1}^{3}P_{j}e^{-i\omega_{j}t} \label{A}\\
Y(t) & = & \sum_{m,n}a_{{s}m}^{\dag}a_{{d}n}e^{i(\epsilon_{{s}m}-\epsilon_{{d}n})t}+a_{{s}m}a_{{d}n}^{\dag}e^{-i(\epsilon_{{s}m}-\epsilon_{{d}n})t}\nonumber\\ \label{Y}
\end{eqnarray}
with
\begin{eqnarray*}
P_{1} & = &-\alpha\beta~\chi_{d}~\sigma_+\\
P_{2} & = &-\alpha\beta~\chi_{d}~\sigma_-\\
P_{3} & = & D-\big(\chi_{1}|e\rangle\langle e|+\chi_{2}|g\rangle\langle g|\big)+\alpha^{2}~\chi_{d}~\varrho_{z},
\end{eqnarray*}
and  $\omega_{1}=-\omega_{2}=-\omega_{0},~\omega_{3}=0$,  $\chi_{d}=\chi_{1}-\chi_{2}$. Here
 $H_{\rm T}(t)$ describes the interaction with the two leads in the Heisenberg picture,
\begin{eqnarray}\label{interaction hamiltonian}
 && H_{{\rm T}}(t)=\sum_{k}\bigg[\Omega_{{l}k}c_{{l}k}e^{-i\epsilon_{{l}k}t}\left(\alpha a_{g}^{\dag}e^{i\omega_{0}t}+\beta a_{e}^{\dag}e^{i\omega_{0}t}\right)\nonumber \\
 && +\Omega_{{r}k}c_{{r}k}^{\dag}e^{i\epsilon_{{r}k}t}\Lambda_{{r}}^{\dag}\left(-\beta a_{g}e^{-i\omega_{0}t}+\alpha a_{e}e^{-i\omega_{0}t}\right)\nonumber \\
&&+ \Omega_{{l}k}^{\ast}c_{{l}k}^{\dag}e^{i\epsilon_{{l}k}t}\left(\alpha a_{g}e^{-i\omega_{0}t}+\beta a_{e}e^{-i\omega_{0}t}\right)\nonumber \\
 && +\Omega_{{r}k}^{\ast}c_{{r}k}e^{-i\epsilon_{{r}k}t}\Lambda_{{r}}\left(-\beta a_{g}^{\dag}e^{i\omega_{0}t}+\alpha a_{e}^{\dag}e^{i\omega_{0}t}\right)\bigg]
\end{eqnarray}
where $a_{g}=|0\rangle\langle g|$ and $a_{e}=|0\rangle\langle e|$ are defined by $d_{1}^{\dag}=|1\rangle\langle0|=\alpha a_{g}^{\dag}+\beta a_{e}^{\dag}$. The above results can also be found in Ref.~\cite{You2010}. We will use them to derive the dynamic model in this paper.


We first consider the interaction of the DQD with the QPC.
Applying  Eq.~(\ref{A}) and  Eq.~(\ref{Y}) to the master equation (\ref{Master equation}), one obtains
\begin{eqnarray}\label{equation}
&&\dot{\rho_s}(t)=\int_0^t{\rm d}t^{\prime}\bigg\{
{\rm Tr}_{{ E}}\left[Y(t^{\prime})\rho_EY(t)\right]\times\nonumber\\
&&\bigg[A(t^{\prime})\rho_s(t)A(t)+A(t)\rho_s(t)A(t^{\prime})\bigg]-{\rm Tr}_{{E}}\left[Y(t)Y(t^{\prime})\rho_E\right]\nonumber\\
&&\bigg[A(t)A(t^{\prime})\rho_s(t)+\rho_sA(t^{\prime})A(t)(t)\bigg]
\bigg\},
\end{eqnarray}
where
\begin{eqnarray}
A(t^{\prime})\rho_sA(t)&=&P_2\rho_sP_1e^{i\omega_0(t-t^{\prime})}+P_1\rho_sP_2e^{i\omega_0(t^{\prime}-t)}\nonumber\\
&&+P_1\rho_sP_1e^{i\omega_0(t+t^{\prime})}+P_2\rho_sP_2e^{-i\omega_0(t+t^{\prime})}\nonumber\\
&&+P_3\rho_sP_1e^{i\omega_0t}+P_3\rho_sP_2e^{-i\omega_0t}\nonumber\\
&&+P_1\rho_sP_3e^{i\omega_0t^{\prime}}+P_2\rho_sP_3e^{-i\omega_0t^{\prime}}\nonumber\\
&&+P_3\rho_sP_3
\end{eqnarray}
and
\begin{eqnarray}
A(t)A(t^{\prime})\rho_s&=&P_1P_2\rho_se^{i\omega_0(t-t^{\prime})}+P_2P_1\rho_se^{i\omega_0(t^{\prime}-t)}\nonumber\\
&&+P_1P_1\rho_se^{i\omega_0(t+t^{\prime})}+P_2P_2\rho_se^{-i\omega_0(t+t^{\prime})}\nonumber\\
&&+P_1P_3\rho_se^{i\omega_0t}+P_2P_3\rho_se^{-i\omega_0t}\nonumber\\
&&+P_3P_1\rho_se^{i\omega_0t^{\prime}}+P_3P_2\rho_se^{-i\omega_0t^{\prime}}\nonumber\\
&&+P_3P_3\rho_s\bigg..
\end{eqnarray}
To simplify the notation, we ignore the subscript $s$ (for system, not source).
Neglecting the fast-oscillating  terms, we obtain the master equation:
\begin{eqnarray}\label{masterequation2}
\dot{\rho}&=&\Gamma_{+}(t)(P_2\rho P_1-P_1P_2\rho)+\Gamma_{-}(t)(P_1\rho P_2-P_2P_1\rho)\nonumber\\
&&+\Gamma_{+}^{\prime}(t)(P_1\rho P_2-\rho P_2P_1)+\Gamma_{-}^{\prime}(t)(P_2\rho P_1-\rho P_1P_2)\nonumber\\
&&+\Lambda(t)(P_3\rho P_3-P_3P_3\rho)+\Lambda^{\prime}(t)(P_3\rho P_3-\rho P_3P_3),\nonumber\\
\end{eqnarray}
where
\begin{eqnarray}
\Gamma_{\pm}(t)&=&\int_0^t\langle Y(s)Y(0)\rangle \exp({\pm i\omega_0s}){\rm d}s\nonumber\\
\Gamma_{\pm}^{\prime}(t)&=&\int_0^t\langle Y(0)Y(s)\rangle \exp({\pm i\omega_0s}){\rm d}s\nonumber\\
\Lambda(t)&=&\int_0^t\langle Y(s)Y(0)\rangle {\rm d}s\nonumber\\
\Lambda^{\prime}(t)&=&\int_0^t\langle Y(0)Y(s)\rangle {\rm d}s.\label{coefficient}
\end{eqnarray}
Here, we define $t-t^{\prime}\equiv s$ and $${\rm Tr}_{{ E}}\{Y(t)Y(t^{\prime})\rho_E\}\equiv\langle Y(s)Y(0)\rangle.$$ Note that the two-time reservoir correlation function $\langle Y(s)Y(0)\rangle$ is
\begin{eqnarray}
\langle Y(s)Y(0)\rangle&=&\sum_{mm^{\prime}nn^{\prime}}\langle Y_{m,n}(s)Y_{m^{\prime},n^{\prime}}(0)\rangle\delta_{m,m^{\prime}}\delta_{n,n^{\prime}}\nonumber\\
&=&\sum_{mn}N_{{s}m}(1-N_{{d}n})e^{i(\epsilon_{{s}m}-\epsilon_{{d}n})s}\nonumber\\
&&+(1-N_{{s}m})N_{{d}n}e^{-i(\epsilon_{{s}m}-\epsilon_{{d}n})s}.
\end{eqnarray}
The notation $\langle~\cdot\cdot\cdot~\rangle$ stands for the statistical average over the source and the drain reservoirs of the QPC. They are assumed to be in thermal equilibrium with the Fermi-Dirac
distribution function

\begin{eqnarray}
N_{{s}m}=\left\{\exp\Big[\frac{(\epsilon_{{s}m}-\mu_{{s}})}{k_{B}T}\Big]+1\right\}^{-1},
\end{eqnarray}
\begin{eqnarray}
N_{dn}=\left\{\exp\Big[\frac{(\epsilon_{dn}-\mu_{d})}{k_{B}T}\Big]+1\right\}^{-1},
\end{eqnarray}
where $\mu_{{s}}=\mu_{F}\pm eV_{{\rm QPC}}/2$ and $\mu_{d}=\mu_{F}\pm eV_{{\rm QPC}}/2$ are the applied QPC bias voltages.
Thus, the coefficients (\ref{coefficient}) become
\begin{eqnarray}
\Gamma_{\pm}(t)&=&\int_0^t{\rm d}s\sum_{mn}\bigg\{N_{sm}(1-N_{dn}) \cos(\epsilon_{{s}m}-\epsilon_{{d}n}\pm\omega_0)s\nonumber\\
&&+(1-N_{sm})N_{dn}\cos(\epsilon_{{s}m}-\epsilon_{{d}n}\mp\omega_0)s\nonumber\\
&&+i\bigg[N_{sm}(1-N_{dn}) \sin(\epsilon_{{s}m}-\epsilon_{{d}n}\pm\omega_0)s\nonumber\\
&&-(1-N_{sm})N_{dn}\sin(\epsilon_{{s}m}-\epsilon_{{d}n}\mp\omega_0)s\bigg]\bigg\},\nonumber\\\label{coefficients5}
\end{eqnarray}
\begin{eqnarray}
\Gamma_{\pm}^{\prime}(t)&=&\int_0^t{\rm d}s\sum_{mn}\bigg\{N_{sm}(1-N_{dn}) \cos(\epsilon_{{s}m}-\epsilon_{{d}n}\mp\omega_0)s\nonumber\\
&&+(1-N_{sm})N_{dn}\cos(\epsilon_{{s}m}-\epsilon_{{d}n}\pm\omega_0)s\nonumber\\
&&+i\bigg[-N_{sm}(1-N_{dn}) \sin(\epsilon_{{s}m}-\epsilon_{{d}n}\mp\omega_0)s\nonumber\\
&&+(1-N_{sm})N_{dn}\sin(\epsilon_{{s}m}-\epsilon_{{d}n}\pm\omega_0)s\bigg]\bigg\},\nonumber\\\label{coefficients6}
\end{eqnarray}
\begin{eqnarray}
\Lambda(t)&=&\int_0^t{\rm d}s\sum_{mn}\bigg\{ i(N_{sm}-N_{dn})\sin(\epsilon_{{s}m}-\epsilon_{{d}n})s\nonumber\\
&&+(N_{sm}+N_{dn}-2N_{sm}N_{dn}) \cos(\epsilon_{{s}m}-\epsilon_{{d}n})s\bigg\},\nonumber\\
\label{coefficients7}
\end{eqnarray}
\begin{eqnarray}
\Lambda^{\prime}(t)&=&\int_0^t{\rm d}s\sum_{mn}\bigg\{- i(N_{sm}-N_{dn})\sin(\epsilon_{{s}m}-\epsilon_{{d}n})s\nonumber\\
&&+(N_{sm}+N_{dn}-2N_{sm}N_{dn}) \cos(\epsilon_{{s}m}-\epsilon_{{d}n})s\bigg\}.\nonumber\\\label{coefficients8}
\end{eqnarray}
For the sake of simplicity,  we define the new variables
\begin{eqnarray}
\omega_{m}\equiv\epsilon_{{s}m}-\mu_{{s}}
\end{eqnarray}
\begin{eqnarray}
\omega_{n}\equiv\epsilon_{{d}n}-\mu_{{d}}.
\end{eqnarray}
It is clear that $$\epsilon_{{s}m}-\epsilon_{{d}n}=\omega_{m}-\omega_{n}+eV_{{\rm QPC}}.$$
Thus, the master equation (\ref{masterequation2}) reduces to
\begin{eqnarray}\label{masterequation3}
\frac{d}{dt}\rho&=&-i\bigg[\eta\Delta_3(t)\sigma_z+\Delta_5(t)P_3P_3, \rho_s\bigg]+\Delta_4(t)\mathcal{D}[P_3]\rho\nonumber\\
&&+\eta\bigg[\Delta_1(t)+\Delta_2(t)\bigg]\mathcal{D}[\sigma_+]\rho\nonumber\\
&&+\eta\bigg[\Delta_1(t)-\Delta_2(t)\bigg]\mathcal{D}[\sigma_-]\rho.
\end{eqnarray}

Now we consider the interaction of the DQD with the two leads.
The Hamiltonian of the two electrodes in the DQD is
\begin{eqnarray}
H_{{\rm Leads}}=\sum_{k}\bigg(\epsilon_{{l}k}c_{{l}k}^{\dag}c_{{{l}k}}+\epsilon_{{r}k}c_{{r}k}^{\dag}c_{{r}k}\bigg),
\end{eqnarray}
where $c_{{l}k}~(c_{{r}k})$ are the annihilation operators of an electron in the left (right) lead with quantum numbers $k$. The tunnelling Hamiltonian of the DQD and the two electrodes is
\begin{eqnarray}
H_{{\rm T}}=\sum_{k}\bigg(\Omega_{{l}k}c_{{l}k}d_{1}^{\dag}+\Omega_{{r}k}c_{{r}k}^{\dag}\Lambda_{{r}}^{\dag}d_{2}\nonumber\\+\Omega_{{l}k}c_{{l}k}^{\dag}d_{1}+\Omega_{{r}k}c_{{r}k}\Lambda_{{r}}d_{2}^{\dag}\bigg),\label{interaction1}
\end{eqnarray}
which depends on the tunnelling coupling strengths $\Omega_{{l}k}$ and $\Omega_{{r}k}$. Here,  $\Lambda_{{r}}^{\dag}$ is the operator used to count the number of electrons that have tunnelled into the right lead. Also,
 $H_{\rm T}(t)$ is the interaction with the two leads in the Heisenberg picture as described in (14).

Similarly, using the interaction Hamiltonian $H_T(t)$ in the master equation (\ref{Master equation}), we obtain the master equation for the system of DQD interacting with the leads (For details, please refer to the Appendix)
\begin{eqnarray}\label{masterequation4}
\frac{d}{dt}\rho=\Gamma_1\mathcal{D}\bigg[\alpha a_g^{\dag}+\beta a_e^{\dag}\bigg]\rho+\Gamma_2\mathcal{D}\bigg[\alpha a_g+\beta a_e\bigg]\rho\nonumber\\
+\Gamma_3\mathcal{D}\bigg[\alpha\Lambda_ra_e^{\dag}+\beta\Lambda_ra_g^{\dag}\bigg]\rho+\Gamma_4\mathcal{D}\bigg[\alpha\Lambda_r^{\dag}a_e+\beta\Lambda_r^{\dag}a_g\bigg]\rho.\nonumber\\
\end{eqnarray}
Now, converting the obtained equation into the Schr\"{o}dinger picture, we have the master equation as follows
\begin{eqnarray}\label{masterequation}
\frac{d}{dt}\rho&=&-i\bigg[H_{\rm DQD}+\eta\Delta_3(t)\sigma_z+\Delta_5(t)P_3P_3, \rho_s\bigg]\nonumber\\
&&+\Delta_4(t)\mathcal{D}[P_3]\rho+\eta\bigg[\Delta_1(t)+\Delta_2(t)\bigg]\mathcal{D}[\sigma_+]\rho\nonumber\\
&&+\eta\bigg[\Delta_1(t)-\Delta_2(t)\bigg]\mathcal{D}[\sigma_-]\rho+\Gamma_1\mathcal{D}\bigg[\alpha a_g^{\dag}+\beta a_e^{\dag}\bigg]\rho\nonumber\\
&&+\Gamma_2\mathcal{D}\bigg[\alpha a_g+\beta a_e\bigg]\rho+\Gamma_3\mathcal{D}\bigg[\alpha\Lambda_ra_e^{\dag}+\beta\Lambda_ra_g^{\dag}\bigg]\rho\nonumber\\
&&+\Gamma_4\mathcal{D}\bigg[\alpha\Lambda_r^{\dag}a_e+\beta\Lambda_r^{\dag}a_g\bigg]\rho,
\end{eqnarray}
where $\eta=\alpha^2\beta^2\chi_d^2$, and the superoperator $\mathcal{D}$
is defined as
\begin{eqnarray}
\mathcal{D}[O]{\rho}=O{\rho}O^{\dag}-\frac{1}{2}(O^{\dag}O{\rho}+{\rho}O^{\dag}O).
\end{eqnarray}
The time-dependent parameters in Eq.~(\ref{masterequation}) are
\begin{eqnarray}\label{Gamma}
\Gamma_1\!\!&=&\!\!\int_0^t\!\!\sum_k\left|\Omega_{l,k}\right|^2\left(\!\!1-\tanh\frac{\epsilon_{l,k}-\mu_l}{2k_BT}\right)\cos(\epsilon_{l,k}-\omega_0)s{\rm d}s\nonumber\\
\Gamma_2\!\!&=&\!\!\int_0^t\!\!\sum_k\left|\Omega_{l,k}\right|^2\left(\!\!1+\tanh\frac{\epsilon_{l,k}-\mu_l}{2k_BT}\right)\cos(\epsilon_{l,k}-\omega_0)s{\rm d}s\nonumber\\
\Gamma_3\!\!&=&\!\!\int_0^t\!\!\sum_k\left|\Omega_{r,k}\right|^2\left(\!\!1-\tanh\frac{\epsilon_{r,k}-\mu_r}{2k_BT}\right)\cos(\epsilon_{r,k}-\omega_0)s{\rm d}s\nonumber\\
\Gamma_4\!\!&=&\!\!\int_0^t\!\!\sum_k\left|\Omega_{r,k}\right|^2\left(\!\!1+\tanh\frac{\epsilon_{r,k}-\mu_r}{2k_BT}\right)\cos(\epsilon_{r,k}-\omega_0)s{\rm d}s,\nonumber\\
\end{eqnarray}
and additional  parameters are given by:
\begin{eqnarray}\label{Delta1}
{\Delta}_{1}(t) & = & \sum_{mn}\Lambda_{1}\int_{0}^{t}{\rm d}s\cos(\omega_{m}-\omega_{n}+eV_{{\rm QPC}})s\cos\omega_{0}s,\nonumber \\
{\Delta}_{2}(t) & = &\sum_{mn}\Lambda_{2} \int_{0}^{t}{\rm d}s\sin(\omega_{m}-\omega_{n}+eV_{{\rm QPC}})s\sin\omega_{0}s, \nonumber\\
{\Delta}_{3}(t) & = &\sum_{mn}\Lambda_{1} \int_{0}^{t}{\rm d}s\cos(\omega_{m}-\omega_{n}+eV_{{\rm QPC}})s\sin\omega_{0}s,\nonumber\\
{\Delta}_{4}(t) & = &4\sum_{mn}\Lambda_{1}\int_{0}^{t}{\rm d}s\cos(\omega_{m}-\omega_{n}+eV_{{\rm QPC}})s,\nonumber \\
{\Delta}_{5}(t) & = & -2\sum_{mn}\Lambda_{2}\int_{0}^{t}{\rm d}s\sin(\omega_{m}-\omega_{n}+eV_{{\rm QPC}})s.
\end{eqnarray}
 The coefficients $\Lambda_{1}$ and $\Lambda_{2}$ are the Fermi functions
\begin{eqnarray}\label{lambda}
\Lambda_{1}(\omega_{m},\omega_{n})=1-\tanh\!\!\left(\frac{\omega_{m}}{2k_{B}T}\right)\tanh\!\!\left(\frac{\omega_{n}}{2k_{B}T}\right),\nonumber\\
\Lambda_{2}(\omega_{m},\omega_{n})=\tanh\!\!\left(\frac{\omega_{m}}{2k_{B}T}\right)-\tanh\!\!\left(\frac{\omega_{n}}{2k_{B}T}\right).
\end{eqnarray}
In the zero temperature limit ($k_BT=0$), the functions $\Lambda_{1}$ and $\Lambda_{2}$ reduce to
\begin{eqnarray}\label{Lambda}
\Lambda_{1}=2\bigg|\Theta(\omega_{m})-\Theta(\omega_{n})\bigg|,\nonumber\\
\Lambda_{2}=2\bigg[\Theta(\omega_{m})-\Theta(\omega_{n})\bigg],
\end{eqnarray}
where $\Theta$ is the Heaviside function \cite{Breuer2002}. Clearly, for different values of $\omega_m$ and $\omega_n$, $\Lambda_1$ and $\Lambda_2$ reduce to different constant values, as shown in Table I.

\centerline{ TABLE I. The Fermi functions $\Lambda_{1}$ and $\Lambda_{2}$.}
\begin{center}
\begin{tabular}{cccc}
\hline\hline \multicolumn{1}{c}{$\omega_m~~~~~~~~~~~~$}
 &\multicolumn{1}{c}{$\omega_n$}
&\multicolumn{1}{c}{$~~~~~~~~~~~~\Lambda_{1}~~~~~~~~~~~~$}
&\multicolumn{1}{c}{$\Lambda_{2}$}\\ \hline
$<0~~~~~~~~~~~~$&$<0$&~~~~~~~~~~~~0 ~~~~~~~~~~~~&0\\

$>0~~~~~~~~~~~~$&$>0$&~~~~~~~~~~~~0~~~~~~~~~~~~ &  0   \\

$>0~~~~~~~~~~~~$&$<0$&~~~~~~~~~~~2~~~~~~~~~~~~& 2
\\

$<0~~~~~~~~~~~~$&$>0$&~~~~~2~~~~~~&$-2$
\\ \hline\hline
\end{tabular}
\end{center}

The master equation obtained in this paper is non-Markovian, but the Born approximation (i.e., second-order perturbation) is still used.  The master equation (\ref{masterequation}) is the result under such an approximation.
This is the first main result of this paper. The advantages of this master equation over previous ones are: (i) it has the Lindblad form and the non-Markovian effects are embodied in the time-dependent coefficients, (ii) the tunnelling and decay effect can be clearly understood in the master equation, and (iii) it can be controlled by manipulating the Hamiltonian.

These results are significant because
two very different kinds of behaviours, Markovian and non-Markovian, are embodied in these time-dependent coefficients (\ref{Delta1}).  In the zero-temperature limit, the Fermi functions (\ref{lambda})  become $0$ or $\pm2$ in certain cases (see Table 1), and the coefficient $\Delta_1(t)-\Delta_2(t)$ of the master equation (\ref{masterequation}) can be reduced to
\begin{eqnarray}
\frac{\Delta_1(t)-\Delta_2(t)}{2}=\sum_{\omega_m>0,~\omega_n<0}\frac{\sin(\omega_m-\omega_n+eV_{\rm QPC}+\omega_0)t}{\omega_m-\omega_n+eV_{\rm QPC}+\omega_0)t}\nonumber\\
+\sum_{\omega_m<0,~\omega_n>0}\frac{\sin(\omega_m-\omega_n+eV_{\rm QPC}-\omega_0)t}{\omega_m-\omega_n+eV_{\rm QPC}-\omega_0)t}.~~~~
\end{eqnarray}
If the  coefficient $\Delta_1(t)-\Delta_2(t)$ is positive, the information is always lost during the time evolution of the DQD quantum system. In certain case, the coefficient becomes negative within certain intervals of time, which displays the non-Markovian behaviour.  Obviously, in these time intervals, the environment compensates some lost information to the quantum system of interest.
The corresponding non-Markovian properties of the negative coefficient were  shown in Fig.~1 of Ref.~\cite{Wei2009}. It is clear that the Markovian dynamics is solely responsible for the long-time limit. We describe the measurement-induced backaction by the tunneling current. We will show that the current is directly determined by the coefficients in Section III. Furthermore, we propose a simple feedback control law to stabilize the current.

\section{Measurement output and induced backaction}
We now investigate  electrons tunnelling in the DQD, and use a feedback control method to stabilize a particular current and reduce the measurement-induced backaction. We consider a zero bias across the DQD, and set $\mu_l=\mu_r=0$.  From  equation (\ref{masterequation}) and the following relations  \cite{You2010,You20132}
\begin{eqnarray}
\langle
n|\Lambda_d^{\dag}\Lambda_d^{}\rho|n\rangle=\rho^{(n)},\nonumber\\
\langle
n|\Lambda_d\Lambda_d^{\dag}\rho|n\rangle=\rho^{(n)},\nonumber\\
\langle
n|\Lambda_d^{\dag}\rho\Lambda_d|n\rangle=\rho^{(n-1)},\nonumber\\
\langle
n|\Lambda_d\rho\Lambda_d^{\dag}|n\rangle=\rho^{(n+1)},
\end{eqnarray}
we obtain a $n$-resolved equation for each reduced density matrix elements:
\begin{eqnarray}\label{resolved equation}
\dot{\rho}_{gg}^{(n)}&=&\eta\bigg(\Delta_1-\Delta_2\bigg)\rho_{gg}^{(n)}-\eta\bigg(\Delta_1+\Delta_2\bigg)\rho_{ee}^{(n)}\nonumber\\
&&+\Gamma_1\alpha^2\rho_{00}^{(n)}-\Gamma_4\beta^2\rho_{gg}^{(n)},\nonumber\\
\dot{\rho}_{ee}^{(n)}&=&\eta\bigg(\Delta_1+\Delta_2\bigg)\rho_{ee}^{(n)}-\eta\bigg(\Delta_1-\Delta_2\bigg)\rho_{gg}^{(n)}\nonumber\\
&&+\Gamma_1\beta^2\rho_{00}^{(n)}-\Gamma_4\alpha^2\rho_{ee}^{(n)},\nonumber\\
\dot{\rho}_{00}^{(n)}&=&-\Gamma_1\rho_{00}^{(n)}+\Gamma_4\bigg(\beta^2\rho_{gg}^{(n-1)}+\alpha^2\rho_{ee}^{(n-1)}\bigg),
\end{eqnarray}
where $n$ is the number of electrons that have tunnelled through the DQD via the right tunnelling barrier at time $t$ so that the density matrix elements $\rho_{{ij}}=\sum_{n}\rho_{ij}^{(n)}~(i,j=0,g,e)$. An electron can tunnel from the left lead into one of the eigenstates $|g\rangle$ and $|e\rangle$ with tunnelling rates $\Gamma_1\alpha^{2}$ and $\Gamma_1\beta^{2}$, respectively. Later, it may tunnel out of the DQD to the right lead with the rates $\Gamma_4\beta^{2}$  from the ground  state or
$\Gamma_4\alpha^{2}$ from the excited  state.

Full counting statistics (FCS) has been widely used to analyze this type of phenomena \cite{Gurvitz1996,Keller2016,Schaller2009}. The FCS method concentrates on the probability distribution for the number
of electron tunneling during a given  period of time. With the help of FCS, one can extract not only the average or variance, but also higher order moments of electron correlations. This method has also been experimentally verified in detecting the higher-order cumulant of electrical noise \cite{Ubbelohde2012, Maisi2014}. Complete information about the transport properties is contained in the probability distribution
$$P(n,t)={\rm Tr}_{s}\{\rho^{(n)}(t)\}=\rho_{00}^{(n)}+\rho_{gg}^{(n)}+\rho_{ee}^{(n)}.$$
 We assume that the time period is much larger than the inverse current frequency. This ensurers that $n\gg1$, on average. The corresponding cumulant generating function (CGF) $\mathcal{F}(\zeta,t)$ is \cite{Braggio2006,Poltl2011}
\begin{eqnarray}
\exp\left[{\mathcal{F}(\zeta,t)}\right]=\sum_{n}P(n,t)\exp(in\zeta),
\end{eqnarray}
where the auxiliary variable $\zeta$ is usually called counting field.
The  power series expansion of the CGF is:
\begin{eqnarray}
\mathcal{F}(\zeta,t)=\sum_{k}C_{k}\frac{i^{k}\zeta^{k}}{k!}.
\end{eqnarray}
It can be proven \cite{Braggio2006} that every moment of $n$ is finite and $\mathcal{F}(\zeta,t)$
is $C^{\infty}$, from which the $k$th order cumulant is \cite{Poltl2011,Braggio2006}
\begin{eqnarray}
C_{k}=\left.\frac{\partial^{k}\mathcal{F}(\zeta,t)}{\partial(i\zeta)^{k}}\right|_{\zeta\rightarrow0}.\label{cumulant1}
\end{eqnarray}
Clearly,
\begin{eqnarray}
F(\zeta,t) & = & \ln\left(\sum_{n}P(n,t)e^{in\zeta}\right)\nonumber \\
 & = & \ln\left(1+\sum_{k=1}^{\infty}\frac{(i\zeta)^{k}}{k!}\left\langle n^{k}\right\rangle \right)\nonumber \\
 & = & \nabla_{1}-\frac{\nabla_{2}}{2}+\frac{\nabla_{3}}{3}-\frac{\nabla_{4}}{4}+\cdots  \nonumber
\end{eqnarray}
where $$\langle n^{k}\rangle=\sum_{n}n^{k}P(n,t).$$ When $$\left|\sum_{n}P(n,t)\left(e^{in\zeta}-1\right)\right|<1,$$
one can use the Taylor expansion to derive the cumulants
\begin{eqnarray}
\nabla_{1} & = & (i\zeta)\left\langle n\right\rangle +\frac{(i\zeta)^{2}}{2!}\left\langle n^{2}\right\rangle +\frac{(i\zeta)^{3}}{3!}\left\langle n^{3}\right\rangle +\nonumber \\
 &  & \frac{(i\zeta)^{4}}{4!}\left\langle n^{4}\right\rangle +\cdots\nonumber \\
\nabla_{2} & = & (i\zeta)^{2}\langle n\rangle^{2}+(i\zeta)^{3}\langle n\rangle\langle n^{2}\rangle+(i\zeta)^{4}\nonumber \\
 &  & \cdot\left(\frac{\langle n^{2}\rangle^{2}}{4}+\frac{2\langle n\rangle\langle n^{3}\rangle}{3!}\right)+\cdots\nonumber \\
\nabla_{3} & = & (i\zeta)^{3}\langle n\rangle^{3}+(i\zeta)^{4}\frac{3\langle n\rangle^{2}\langle n^{2}\rangle}{2}+\cdots\nonumber \\
\nabla_{4} & = & (i\zeta)^{4}\langle n\rangle^{4}+\cdots \nonumber
\end{eqnarray}

\begin{figure}
\setlength{\abovecaptionskip}{6pt}
\centerline{\scalebox{0.48}[0.55]{\includegraphics{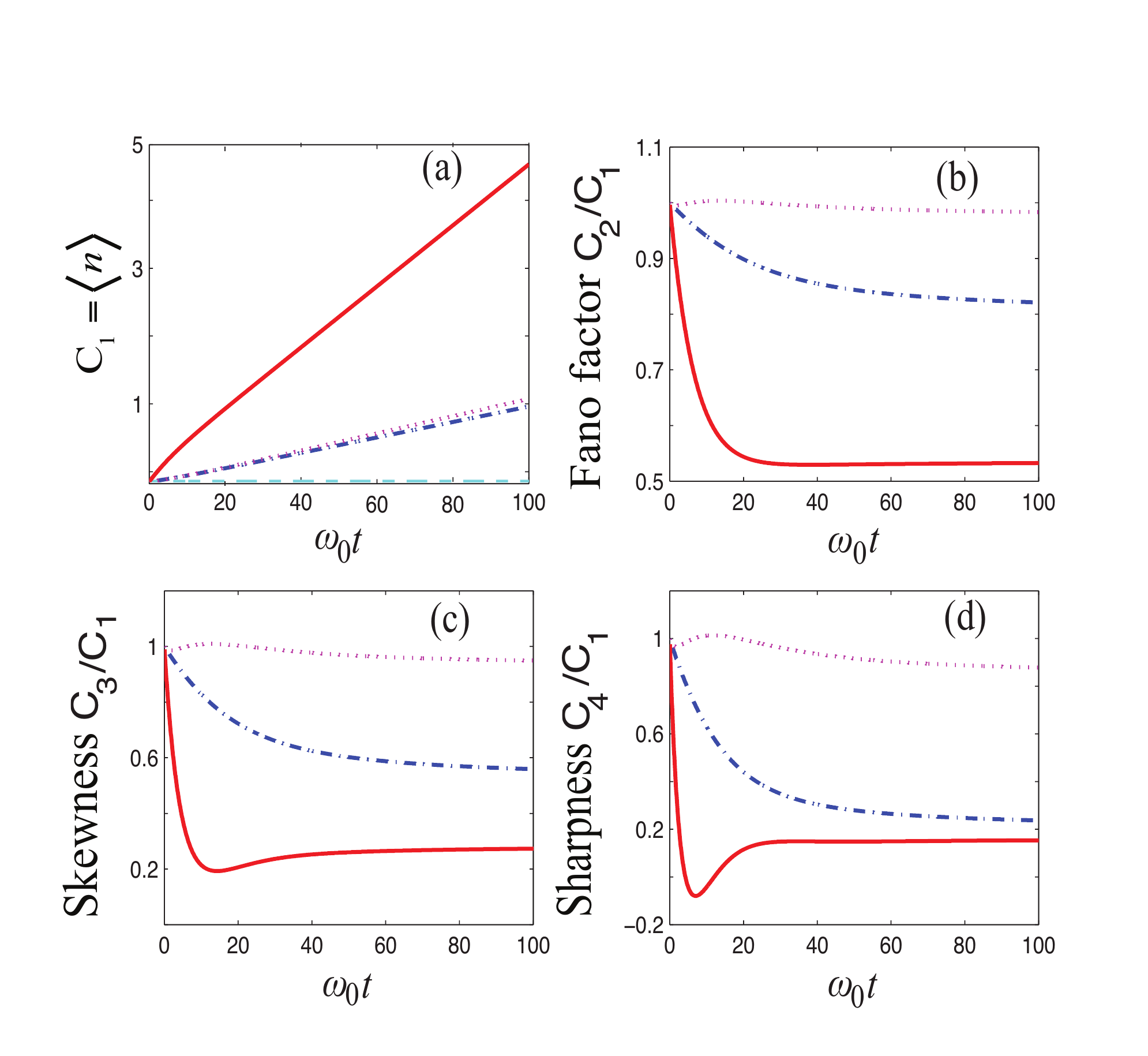}}}
\caption{The mean transport density $C_{1}=\langle n\rangle$, the Fano factor $C_{2}/C_{1}$, the normalized skewness $C_{3}/C_{1}$, and the normalized sharpness $C_{4}/C_{1}$ versus the detection time
for various electron tunneling rates. Here, the cyan-dash lines denote $\Gamma_{4}=0$, the blue-dash-dot lines denote $\Gamma_{1}=0.1,\Gamma_{4}=0.1$,
the purple-dot lines denote $\Gamma_{1}=0.5,\Gamma_{4}=0.1$, and the red-solid lines denote $\Gamma_{1}=0.1,\Gamma_{4}=0.5$.}
\end{figure}
From Eq.~(\ref{cumulant1}), the first four  cumulants of the generating function can be expressed as
\begin{eqnarray} \label{cumulant}
C_{1}(t) & = & \langle n\rangle\\
C_{2}(t) & = & \langle n^{2}\rangle-\langle n\rangle^{2}\nonumber \\
C_{3}(t) & = & \langle n^{3}\rangle-3\langle n\rangle\langle n^{2}\rangle+2\langle n\rangle^{3}\nonumber \\
C_{4}(t) & = & \langle n^{4}\rangle-3\langle n^{2}\rangle^{2}-4\langle n\rangle\langle n^{3}\rangle+12\langle n\rangle^{2}\langle n^{2}\rangle-6\langle n\rangle^{4}.\nonumber
\end{eqnarray}
They denote the mean, variance, asymmetry (``skewness'') and kurtosis
(``sharpness''), respectively.
The mean transport density $C_{1}(t)$, the Fano factor
$C_{2}/C_{1}$, the normalized skewness $C_{3}/C_{1}$, and the normalized sharpness $C_{4}/C_{1}$ are plotted in Fig. 2 for various electron tunneling rates. Here, the cyan-dash lines denote $\Gamma_{4}=0$, the blue-dash-dot lines denote $\Gamma_{1}=0.1,\Gamma_{4}=0.1$, the purple-dot lines denote $\Gamma_{1}=0.5,\Gamma_{4}=0.1$, and the red-solid lines denote $\Gamma_{1}=0.1,\Gamma_{4}=0.5$.
The coefficients are set as $\chi_{d}=0.5,\epsilon=108\ \mu eV,\Omega=32\ \mu eV$, $V_{{\rm QPC}}=400\ \mu eV$,
and the initial condition of Eq. (\ref{resolved equation}) is $\rho_{gg}^{(n)}(0)=\delta_{n,0}$.
From Fig. 2(a), we find that if the tunneling rate
$\Gamma_{4}=0$, the cumulants will keep zero.
If we set $\Gamma_{4}=0.1$
and change $\Gamma_{1}$ from $0.1$ to $0.5$, there are no significant
changes. However, when the tunneling rate of the
drain is enhanced to $0.5$, drastic changes occur to all the cumulants.
The tunnelling current can be  extracted from the mean $C_{1}$ in Eq. (\ref{cumulant}),
\begin{eqnarray}
I(t)&=&e\frac{d}{dt}\langle n\rangle=e\sum_{n}n\dot{P}_{n}\nonumber \\
 & = & e\Gamma_{{4}}\sum_{n}n\left[\beta^{2}(\rho_{gg}^{(n-1)}-\rho_{gg}^{(n)})+\alpha^{2}(\rho_{ee}^{(n-1)}-\rho_{ee}^{(n)})\right]\nonumber \\
 & = & e\Gamma_{{4}}\left(\beta^{2}\rho_{gg}(t)+\alpha^{2}\rho_{ee}(t)\right).
\end{eqnarray}
This series is cut-off with $\rho_{gg}^{(-1)}=0$ and $\rho_{ee}^{(-1)}=0$, and normalization holds as $$\sum_{n}\rho_{gg}^{(n)}=\rho_{gg}$$ and $$\sum_{n}\rho_{ee}^{(n)}=\rho_{ee}.$$ It is clear that the current linearly depends  on the electron tunneling rate $\Gamma_4$ and the population of the quantum state.

\begin{figure}[h!]
\setlength{\abovecaptionskip}{6pt}
\centerline{\scalebox{0.62}[0.62]{\includegraphics{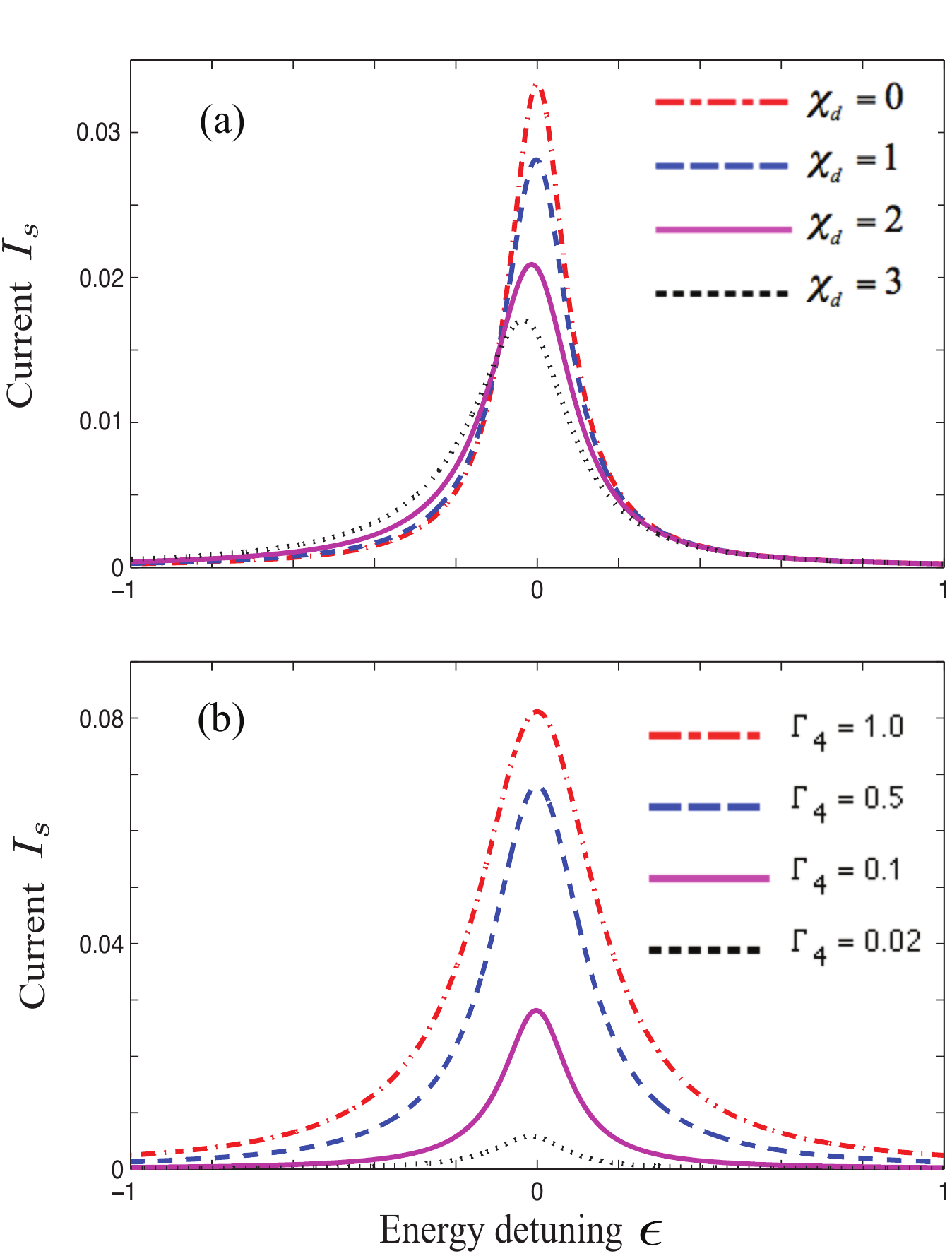}}}
\caption{ Stationary transport current $I_{s}$ as a function of the energy detuning for different parameters.
Fig. 3(a) demonstrates that by increasing the coupling strength $\chi_{d}$, the current resonance becomes asymmetrically broadened a bit and its maximum values decreases; Fig. 3(b) indicates that the current is more significant for a larger tunneling rate $\Gamma_{4}$.}
\end{figure}


In the long-time limit, the stationary current $I_{s}=I(t\rightarrow\infty)$ reads
\begin{eqnarray}\label{stationary current}
I_{s}=\frac{e\Gamma_{{1}}\Gamma_{{4}}(2\beta^{2}\chi_{{\rm d}}^{2}\Delta_{2}+\Gamma_4)}{I_1+I_2},\end{eqnarray}
where
\begin{eqnarray}
I_1&=&\Gamma_{{1}}\Gamma_{{4}}\left(\frac{\alpha^{2}}{\beta^{2}}+\frac{\beta^{2}}{\alpha^{2}}\right)+\Gamma_{{4}}^{2}-\chi_{{\rm d}}^{2}\Gamma_{{4}}\Delta_{2}(2\alpha^{2}-1),\nonumber\\
I_2&=&\chi_{{\rm d}}^{2}(\Gamma_4(\Delta_{1}+2\Delta_{2})+2\Gamma_{4}\Delta_{1}).\nonumber
\end{eqnarray}
It is clear that the stationary current is determined by the tunneling rates [$\Gamma_1$ and $\Gamma_4$, defined in Eq. (\ref{Gamma})], the measurement-induced decay rates [$\Delta_{1}$ and $\Delta_{2}$, defined in Eq. (\ref{Delta1})] and the coupling strength $\chi_{d}^{2}$. In Fig. 3, we plot the stationary current as a function of the  energy detuning $\epsilon$ for various  values of the coupling strength $\chi_d^2$ and tunnelling rates $\Gamma_4$.
In Fig. 3(a), by increasing the coupling strength $\chi_{d}$, the current resonance becomes asymmetrically broadened a bit and its maximum value decreases. This is because the measurement-induced backaction becomes stronger when $\chi_{d}$  is larger. From Fig. 3(b), we can also find that the current is more significant for a larger tunnelling rate $\Gamma_{4}$. The reason is that the larger the tunnelling rate the more electrons can tunnel. In this paper, we investigate the backaction by considering the influence of the energy detuning. It is worth noting that recently Ref. \cite{You20132} demonstrated  considerable QPC-induced backaction by investigating the influence of the QPC parameter $eV_{\rm QPC}$, which is embodied in the decay coefficients of the master equation (\ref{masterequation}). The authors in \cite{You20132} showed that when the QPC bias energy $eV_{\rm QPC}$ goes beyond a constant value, the current increases with the magnitude of the voltage applied across the QPC in a nearly linear manner.
 Here, the energy detuning influence is embedded in the Hamiltonian in the master equation (\ref{masterequation}). Within this particular structure, the quantum Hamiltonian control method \cite{FT} may be used to stabilize a target quantum state and optimize the resulting convergence speed. Below, we use this method to stabilize the tunneling current.

\section{Stabilization of the tunneling current}
In this section we specifically study how to stabilize a particular tunneling current. This is similar to tracking control \cite{Xue:2016}  in classical control theory. Feedback is an essential concept in control theory, for its prominent capability in dealing with various uncertainties. In general, quantum semi-classical feedback control consists of applying a conditional Hamiltonian to a quantum system to obtain  desired results, i.e., engineering the system evolution in the form
\begin{eqnarray}
-i\sum_{j}u_{j}(t)[H_{j},~\rho].
\end{eqnarray}
The control law $u_{j}(t)$ is a function of the measurement output \begin{eqnarray}u_{j}(t)=f(t,I(t)).\end{eqnarray}
Direct quantum feedback with no time delay is possible if the control is sufficiently strong and fast.
In order to obtain a particular tunnelling current, here we design a feedback control law to the double quantum dot. From Eq.~(\ref{Ham}) there are two adjustable parameters that are experimentally accessible,i.e.,  the energy level $\epsilon$ and the coupling $\Omega$ in the system Hamiltonian.  We assume that the pulses are ideal and act as an instantaneous unitary operation on the system, so that
no coupling to the surroundings needs to be considered during this process. The corresponding system Hamiltonian  becomes
\begin{eqnarray}
H_{{\rm DQD}}^{\prime}=\frac{\epsilon+u_{1}(t)}{2}\sigma_{z}+\Big[\Omega+u_{2}(t)\Big]\sigma_{x}.\label{control}
\end{eqnarray}
Explicitly, $u_1(t)$ can be achieved by applying the time-dependent gate voltages on the two dots to vary the level difference between them, and $u_2(t)$ can be achieved by applying a time-dependent gate voltage  between the two dots to vary the interdot barrier.
Based on the above analysis,  we propose a simple feedback control law to realize and stabilize a particular tunnelling current $I_{0}$
\begin{eqnarray}
u_{i}(t)={\rm sgn}\big[I_{0}-I(t)\big]\eta_{i}\exp\Big(-\frac{1}{k|I_{0}-I(t)|}\Big),\label{controller}
\end{eqnarray}
where $\eta_{i}>0$ is the control amplitude, $k>0$ is an adaptive factor, $I_{0}$ is the target electron current, and ${\rm sgn}$ is the sign function,
\begin{equation}
\label{character}{\rm sgn}(x)=\left\{
 \begin{array}{rcl}
\!\!\!&1& ~~~{\rm if}~~~x>0,\\
\!\!\!&0&~~~{\rm if}~~~x=0,\\
\!\!\!&-1&~~~{\rm if}~~~x<0.\\
\end{array}\right.
 \end{equation}
Thus, the tunneling current $I(t)$ in Eq.~(\ref{controller}) determines whether to increase or decrease the tunnelling coupling and the energy level (speed up or slow down the electron tunnelling).  It is clear that the controller (\ref{controller}) has the characteristics
\begin{figure}
\setlength{\abovecaptionskip}{6pt}
\centerline{\scalebox{0.65}[0.65]{\includegraphics{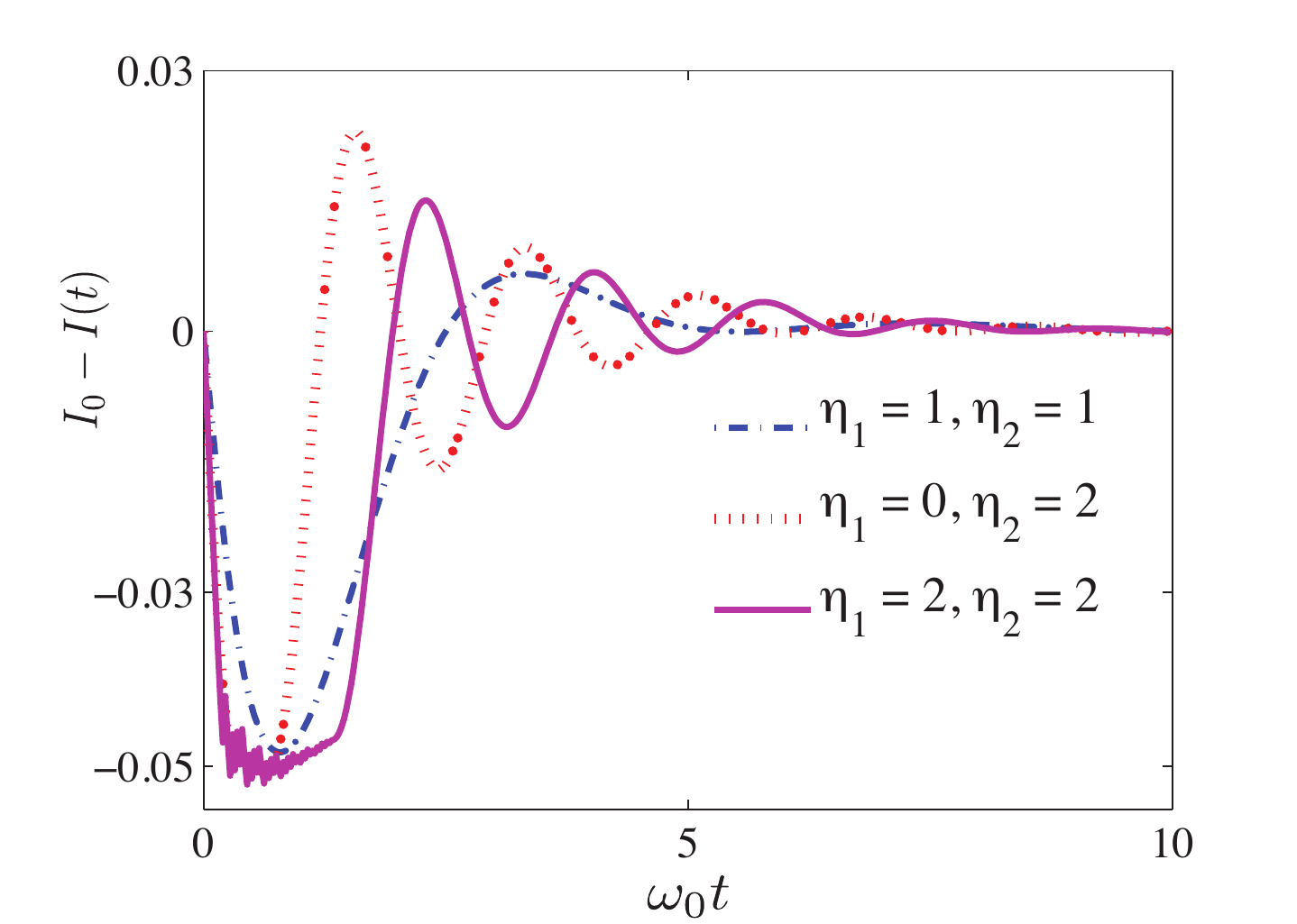}}}
\caption{ Current difference, $I_{0}-I(t)$ as a function of $\omega_0t$, for various control amplitudes. The blue dot-dashed line corresponds to $\eta_1=\eta_2=1$, the red dotted line corresponds to $\eta_1=0,~\eta_2=2$,  and the purple solid line for $\eta_1=\eta_2=2$. Note that the feedback control law can quickly drive the current $I(t)$  to the expect one $I_0$. }
\end{figure}

\begin{equation}
\label{character}\left\{
 \begin{array}{rcl}
\!\!\!\!&&\lim_{I(t)\rightarrow I_{0}}\ u_{i}(t) \ =\   0,\\
\!\!\!\!&&u_{i}(t) \to -\eta_{i},\ \ \ \ \ \textrm{if}\ \ I(t)\gg I_{0},\\
\!\!\!\!&&u_{i}(t) \to +\eta_{i},\ \ \ \ \ \textrm{if}\ \ I(t)\ll I_{0}.
\end{array}\right.
 \end{equation}
Equation (\ref{character}) demonstrates that if the current $I(t)$ is smaller than the target one, a stronger coupling strength is applied, which will speed up the electron transport. If the current $I(t)$ is larger than the target one, a weaker coupling strength is applied, which will slow down the electron tunnelling. This algorithm (feedback control process) will end when $I(t)\to I_{0}$.

In Fig.~4 we show the numerical results for  the current difference $I_{0}-I(t)$ as a function of $\omega_0t$ for various control amplitudes. Here we set $I_{0}=0.1$ and the factor $k=5\times10^4$. Note that all the parameters are dimensionless. It is clear that with the feedback control law,  the current $I(t)$ quickly convergences to the target value $I_{0}$.  It is also demonstrated in Fig. 5 that for various control amplitudes $\eta_i$, the behaviours of the feedback controller are quite different.
Usually, the larger the amplitude, the more oscillation behaviour. The reason is that the sign function  determines whether to increase or decrease the tunnelling coupling and the energy level,  the amplitude indicates how large the control law is and the factor indicates how quick the current $I(t)$ convergences to the target value. We consider that the feedback control (30) acts on an ensemble and   the measurement is implemented over many identical systems simultaneously. Thanks to the current technology, the measurement and feedback control of a charge qubit in a DQD can be implemented in the experiment; e.g., see \cite{Cao:2012}.

\section{Conclusion}
We considered the non-Markovian dynamics of open quantum systems and a time-dependent master equation for a DQD interacting with the QPC was derived. We further illustrated several basic features of quantum non-Markovianity by analyzing the first four cumulants.
Moreover, we  propose a control algorithm to realize and stabilize a particular tunnelling current
from its surrounding environment. The strategy can be summarized as follows: we calculate the cumulants of full counting statistics by the  master equation, and then deduce the tunnelling current.  Based on the tunnelling current, a Hamiltonian feedback control law was proposed to realize and stabilize a particular measurement current. We considered a specific physical set-up of a double quantum dot measured by a nearby quantum point contact. Theoretical analysis and numerical simulations show that the control algorithm can make the current quickly convergent to the target value, and thus enhance the performance of the quantum measurement.


%

\appendices
\section*{Appendix: Master equation for DQD interacting with the leads}
Using the interaction Hamiltonian $H_T(t)$ in the master equation (\ref{Master equation}), we obtain the master equation for the system of DQD interacting with the leads as follows \cite{CuiCCC}:
\begin{eqnarray}\label{master2}
\dot{\rho}=\alpha^2{\rm I}+\beta^2{\rm II}+\beta^2{\rm III}+\alpha^2{\rm IV},
\end{eqnarray}
 where
\begin{eqnarray}
{\rm I}&=&\Gamma_{l,-}\big(a_g^{\dag}\rho a_g-a_ga_g^{\dag}\rho\big)+\Gamma^{\prime}_{l,-}\big(a_g\rho a_g^{\dag}-\rho a_g^{\dag}a_g\big)\nonumber\\
&&+\Gamma_{l,+}\big(a_g\rho a_g^{\dag}-a_g^{\dag}a_g\rho \big)+\Gamma^{\prime}_{l,+}\big(a_g^{\dag}\rho a_g-\rho a_ga_g^{\dag}\big)\nonumber\\
&=&-i\left[\frac{\Gamma_{l,-}-\Gamma_{l,+}^{\prime}}{2i}a_ga_g^{\dag}-\frac{\Gamma_{l,+}-\Gamma_{l,-}^{\prime}}{2i}a_g^{\dag}a_g, \rho\right]\nonumber\\
&&+(\Gamma_{l,-}+\Gamma^{\prime}_{l,+})\mathcal{D}[a_g^{\dag}]{\rho}+(\Gamma_{l,+}+\Gamma^{\prime}_{l,-})\mathcal{D}[a_g]{\rho},\nonumber\\
\end{eqnarray}

\begin{eqnarray}
{\rm II}&=&\Gamma_{l,-}\big(a_e^{\dag}\rho a_e-a_ea_e^{\dag}\rho \big)+\Gamma^{\prime}_{l,-}\big(a_e\rho a_e^{\dag}-\rho a_e^{\dag}a_e\big)\nonumber\\
&&+\Gamma_{l,+}\big(a_e\rho a_e^{\dag}-a_e^{\dag}e_g\rho \big)+\Gamma^{\prime}_{l,+}\big(a_e^{\dag}\rho a_e-\rho a_ea_e^{\dag}\big)\nonumber\\
&=&-i\left[\frac{\Gamma_{l,-}-\Gamma_{l,+}^{\prime}}{2i}a_ea_e^{\dag}-\frac{\Gamma_{l,+}-\Gamma_{l,-}^{\prime}}{2i}a_e^{\dag}a_e, \rho \right]\nonumber\\
&&+(\Gamma_{l,-}+\Gamma^{\prime}_{l,+})\mathcal{D}[a_e^{\dag}]{\rho}+(\Gamma_{l,+}+\Gamma^{\prime}_{l,-})\mathcal{D}[a_e]{\rho},\nonumber\\
\end{eqnarray}

\begin{eqnarray}
{\rm III}&=&\Gamma_{r,-}\big(\Lambda_ra_g^{\dag}\rho\Lambda_r^{\dag}a_g-\Lambda_r^{\dag}a_g\Lambda_ra_g^{\dag}\rho\big)\nonumber\\
&&+\Gamma^{\prime}_{r,-}\big(\Lambda_r^{\dag}a_g\rho\Lambda_r a_g^{\dag}-\rho\Lambda_ra_g^{\dag}\Lambda_r^{\dag}a_g\big)\nonumber\\
&&+\Gamma_{r,+}\big(\Lambda_r^{\dag}a_g\rho\Lambda_ra_g^{\dag}-\Lambda_ra_g^{\dag}\Lambda_r^{\dag}a_g\rho\big)\nonumber\\
&&+\Gamma^{\prime}_{r,+}\big(\Lambda_ra_g^{\dag}\rho\Lambda_r^{\dag}a_g-\rho\Lambda_r^{\dag}a_g\Lambda_ra_g^{\dag}\big)\nonumber\\
&=&(\Gamma_{r,-}+\Gamma^{\prime}_{r,+})\mathcal{D}[\Lambda_ra_g^{\dag}]{\rho}+(\Gamma_{r,+}+\Gamma^{\prime}_{r,-})\mathcal{D}[\Lambda_r^{\dag}a_g]{\rho}\nonumber\\
&&-i\left[\frac{\Gamma_{r,-}-\Gamma_{r,+}^{\prime}}{2i}\Lambda_r^{\dag}a_g\Lambda_ra_g^{\dag}, ~\rho \right]\nonumber\\
&&-i\left[-\frac{\Gamma_{r,+}-\Gamma_{r,-}^{\prime}}{2i}\Lambda_ra_g^{\dag}\Lambda_r^{\dag}a_g,~ \rho\right],\nonumber\\
\end{eqnarray}

\begin{eqnarray}
{\rm IV}&=&\Gamma_{r,-}\big(\Lambda_ra_e^{\dag}\rho\Lambda_r^{\dag}a_e-\Lambda_r^{\dag}a_e\Lambda_ra_e^{\dag}\rho\big)\nonumber\\
&&+\Gamma^{\prime}_{r,-}\big(\Lambda_r^{\dag}a_e\rho\Lambda_r a_e^{\dag}-\rho\Lambda_ra_e^{\dag}\Lambda_r^{\dag}a_e\big)\nonumber\\
&&+\Gamma_{r,+}\big(\Lambda_r^{\dag}a_e\rho\Lambda_ra_e^{\dag}-\Lambda_ra_e^{\dag}\Lambda_r^{\dag}a_e\rho\big)\nonumber\\
&&+\Gamma^{\prime}_{r,+}\big(\Lambda_ra_e^{\dag}\rho\Lambda_r^{\dag}a_e-\rho\Lambda_r^{\dag}a_e\Lambda_ra_e^{\dag}\big)\nonumber\\
&=&(\Gamma_{r,-}+\Gamma^{\prime}_{r,+})\mathcal{D}[\Lambda_ra_e^{\dag}]{\rho}+(\Gamma_{r,+}+\Gamma^{\prime}_{r,-})\mathcal{D}[\Lambda_r^{\dag}a_e]{\rho}.\nonumber\\
&&-i\left[\frac{\Gamma_{r,-}-\Gamma_{r,+}^{\prime}}{2i}\Lambda_r^{\dag}a_e\Lambda_ra_e^{\dag},~ \rho \right]\nonumber\\
&&-i\left[-\frac{\Gamma_{r,+}-\Gamma_{r,-}^{\prime}}{2i}\Lambda_ra_e^{\dag}\Lambda_r^{\dag}a_e, \rho \right].\nonumber\\
\end{eqnarray}
The coefficients are
\begin{eqnarray}
\Gamma_{l,-}=\int_0^t\sum_k\left|\Omega_{l,k}\right|^2\left \langle c_{l,k}^{\dag}c_{l,k}\right\rangle e^{i(\epsilon_{l,k}-\omega_0)s}{\rm d}s,\nonumber\\
\Gamma_{l,+}=\int_0^t\sum_k\left|\Omega_{l,k}\right|^2\left\langle c_{l,k}c_{l(r),k}^{\dag}\right\rangle e^{-i(\epsilon_{l,k}-\omega_0)s} {\rm d}s,\nonumber\\
\Gamma_{l,-}^{\prime}=\int_0^t\sum_k\left|\Omega_{l,k}\right|^2 \left\langle c_{l,k}c_{l,k}^{\dag}\right\rangle e^{i(\epsilon_{l,k}-\omega_0)s} {\rm d}s,\nonumber\\
\Gamma_{l,+}^{\prime}=\int_0^t\sum_k\left|\Omega_{l,k}\right|^2 \left\langle c_{l,k}^{\dag}c_{l,k}\right\rangle e^{-i(\epsilon_{l,k}-\omega_0)s}{\rm d}s,\nonumber\\
\Gamma_{r,-}=\int_0^t\sum_k\left|\Omega_{r,k}\right|^2\left \langle c_{r,k}^{\dag}c_{r,k}\right\rangle e^{i(\epsilon_{r,k}-\omega_0)s}{\rm d}s,\nonumber\\
\Gamma_{r,+}=\int_0^t\sum_k\left|\Omega_{r,k}\right|^2\left\langle c_{r,k}c_{r,k}^{\dag}\right\rangle e^{-i(\epsilon_{r,k}-\omega_0)s} {\rm d}s,\nonumber\\
\Gamma_{r,-}^{\prime}=\int_0^t\sum_k\left|\Omega_{r,k}\right|^2 \left\langle c_{r,k}c_{r,k}^{\dag}\right\rangle e^{i(\epsilon_{r,k}-\omega_0)s} {\rm d}s,\nonumber\\
\Gamma_{r,+}^{\prime}=\int_0^t\sum_k\left|\Omega_{r,k}\right|^2 \left\langle c_{r,k}^{\dag}c_{r,k}\right\rangle e^{-i(\epsilon_{r,k}-\omega_0)s}{\rm d}s,\nonumber
\end{eqnarray}
with the Fermi-Dirac distribution function
\begin{eqnarray}
\left \langle c_{l,k}^{\dag}c_{l,k}\right\rangle&=&\left[{e^{\left(\epsilon_{l,k}-\mu_{l}\right)/k_BT}+1}\right]^{-1}\nonumber\\
&=&\frac{1}{2}\left[1-\tanh\frac{\epsilon_{l,k}-\mu_{l}}{2k_BT}\right]
\end{eqnarray}
and
\begin{eqnarray}
\left \langle c_{r,k}^{\dag}c_{r,k}\right\rangle&=&\left[{e^{\left(\epsilon_{r,k}-\mu_{r}\right)/k_BT}+1}\right]^{-1}\nonumber\\
&=&\frac{1}{2}\left[1-\tanh\frac{\epsilon_{r,k}-\mu_{r}}{2k_BT}\right].
\end{eqnarray}
Thus, the master equation (\ref{master2}) reduces to
\begin{eqnarray}\label{masterequation4}
\frac{d}{dt}\rho=\Gamma_1\mathcal{D}\bigg[\alpha a_g^{\dag}+\beta a_e^{\dag}\bigg]\rho+\Gamma_2\mathcal{D}\bigg[\alpha a_g+\beta a_e\bigg]\rho\nonumber\\
+\Gamma_3\mathcal{D}\bigg[\alpha\Lambda_ra_e^{\dag}+\beta\Lambda_ra_g^{\dag}\bigg]\rho+\Gamma_4\mathcal{D}\bigg[\alpha\Lambda_r^{\dag}a_e+\beta\Lambda_r^{\dag}a_g\bigg]\rho,\nonumber\\
\end{eqnarray}
where
\begin{eqnarray}
&&\Gamma_1=\Gamma_{l,-}+\Gamma_{l,+}^{\prime}\nonumber\\
&=&\int_0^t\sum_k\left|\Omega_{l,k}\right|^2\left(1-\tanh\frac{\epsilon_{l,k}-\mu_l}{2k_BT}\right)\cos(\epsilon_{l,k}-\omega_0)s{\rm d}s\nonumber\\
&&\Gamma_2=\Gamma_{l,+}+\Gamma_{l,-}^{\prime}\nonumber\\
&=&\int_0^t\sum_k\left|\Omega_{l,k}\right|^2\left(1+\tanh\frac{\epsilon_{l,k}-\mu_l}{2k_BT}\right)\cos(\epsilon_{l,k}-\omega_0)s{\rm d}s\nonumber\\
&&\Gamma_3=\Gamma_{r,-}+\Gamma_{r,+}^{\prime}\nonumber\\
&=&\int_0^t\sum_k\left|\Omega_{r,k}\right|^2\left(1-\tanh\frac{\epsilon_{r,k}-\mu_r}{2k_BT}\right)\cos(\epsilon_{r,k}-\omega_0)s{\rm d}s\nonumber\\
&&\Gamma_4=\Gamma_{r,+}+\Gamma_{r,-}^{\prime}\nonumber\\
&=&\int_0^t\sum_k\left|\Omega_{r,k}\right|^2\left(1+\tanh\frac{\epsilon_{r,k}-\mu_r}{2k_BT}\right)\cos(\epsilon_{r,k}-\omega_0)s{\rm d}s.\nonumber\\
\end{eqnarray}

\section*{Acknowledgment}
The authors would like to thank
 Prof. Franco Nori and Prof. Jianqiang You for helpful comments.

\ifCLASSOPTIONcaptionsoff
  \newpage
\fi

\end{document}